# Ultrafast Internal Conversion in Ethylene. I. The Excited State Lifetime

H. Tao,[1,2] T. K. Allison,[3,4] T. W. Wright,[4,5] A. M. Stooke,[3,4] C. Khurmi,[4] J. van Tilborg,[4] Y. Liu,[3,4] R. W. Falcone,[3,4] A. Belkacem,[4*] and T. J. Martinez[1,2*]

[1]*Dept of Chemistry and PULSE Institute, Stanford University, Stanford, CA 94305*
[2]*SLAC National Accelerator Laboratory, Menlo Park, CA 94309*
[3]*Dept of Physics, University of California at Berkeley, Berkeley, CA 94720 and*
[4]*Ultrafast X-ray Science Laboratory, Lawrence Berkeley National Lab, Berkeley, CA 94720*
[5]*Dept of Chemistry, University of California at Davis, Davis, CA 95616*

**Abstract**

Using a combined theoretical and experimental approach, we investigate the non-adiabatic dynamics of the prototypical ethylene ($C_2H_4$) molecule upon $\pi \rightarrow \pi^*$ excitation. In this first part of a two part series, we focus on the lifetime of the excited electronic state. The femtosecond Time-Resolved Photoelectron Spectrum (TRPES) of ethylene is simulated based on our recent molecular dynamics simulation using the *ab initio* multiple spawning method (AIMS) with Multi-State Second Order Perturbation Theory (Tao, *et al*. J. Phys. Chem. A **113** 13656 2009). We find excellent agreement between the TRPES calculation and the photoion signal observed in a pump-probe experiment using femtosecond vacuum ultraviolet (hv = 7.7 eV) pulses for both pump and probe. These results explain the apparent discrepancy over the excited state lifetime between theory and experiment that has existed for ten years, with experiments (e.g., Farmanara, *et al*. Chem. Phys. Lett. **288** 518 1998 and Kosma, *et al*. J. Phys. Chem. A **112** 7514 2008) reporting much shorter lifetimes than predicted by theory. Investigation of the TRPES indicates that the fast decay of the photoion yield originates from both energetic and electronic factors, with the energetic factor playing a larger role in shaping the signal.

* To whom correspondence should be addressed:

E-mail: Todd.Martinez@stanford.edu

E-mail: abelkacem@lbl.gov

## Introduction

Photochemical reactions are of fundamental importance in nature. For example, the ultrafast isomerization of the excited rhodopsin molecule is the crucial first step in vision,[1,2] the photoinduced ring-opening reaction of cyclohexadiene represents the first step in a number of important biological processes,[3-5] and the quenching of electronic excitation in DNA bases is responsible for their ultraviolet photostability.[6] At least one general trend emerges from these systems - conical intersections (CIs) facilitate ultrafast electronic relaxation and the efficient conversion of electronic excitation into vibrational energy. The importance, complexity and relatively fast nature of these reactions make them popular targets of accurate molecular dynamics studies. However, despite the increasing sophistication of dynamics simulations, there remain discrepancies between experiments and theory.

Ethylene serves as an excellent example here. Although it is the simplest molecule with a carbon double bond, it exhibits rich internal conversion dynamics. Thus, ethylene has attracted an enormous amount of attention from both experimentalists and theoreticians. Various theoretical methods predict that after $\pi \rightarrow \pi^*$ excitation, the molecule experiences an ultrafast decay back to the ground state through two general classes of CIs, one occurring at twisted-pyramidalized structures and the other near ethylidene-like configurations ($CH_3CH$) where one of the hydrogens has migrated across the double bond.[7-10] Time-resolved measurements have found the timescale of this process to be approximately 50 femtoseconds (fs).[11-15] Dynamics simulations, on the other hand, predict a much longer excited state lifetime,[7,16-19] in the range of 89 to 180 fs. As the accuracy of electronic structure implemented in dynamics simulations increases, the calculated excited state lifetime tends to shorten, approaching the experimental value. For example, our recent simulation using the *ab initio* multiple spawning method (AIMS) with Multi-State Second Order Perturbation Theory (MS-CASPT2) predicted 89 fs for the excited state lifetime. However, this still lies well outside the confidence intervals proposed by the experiments.[11-15] In this paper we focus on the lifetime of the electronic excited state and the resolution of this apparent discrepancy. In part 2 of this series, we will report on the experimental observation of pathways quenching through the two different classes of conical intersections and the branching ratio between these pathways.

There are two aspects of previous measurements that could complicate the comparison between experiment and theory. First, some of the previous time resolved measurements[11,14] initiated the excited state dynamics using a pump pulse with a carrier frequency more than 1 eV to the red of the absorption maximum (~7.66 eV). Therefore, the center of the nascent excited state wavefunction is displaced from the Franck-Condon point. In the case of ethylene, this displacement would likely be along the torsional coordinate that leads to the conical intersections (because torsion about the C=C bond lowers the energy gap between $S_0$ and the π→π* excited state). This would lead to an observed shortening of the excited state lifetime, since one is effectively starting in a region that is closer to the intersection seam than the planar Franck-Condon geometry. Secondly, many of the previous experiments[11,14,15] have used intense long-wavelength pulses to ionize ethylene from the excited state surface via multiphoton absorption. The multiphoton nature of the probe pulse complicates direct comparisons between theory and experiment. Numerical simulation of such an experiment must not only accurately calculate the multi-surface dynamics initiated by the pump pulse, but also the complex multiphoton ionization process of the probe, including the multiphoton absorption cross-sections in the excited state. From the theoretical perspective, the ideal experiment would avoid both of these issues, exciting the molecule with a pump pulse centered on the absorption maximum and probing with a high-energy photon at low intensities where ionization was induced by single photon absorption (which simplifies the calculation of the required photoionization cross-sections). In the present work, we carried out Time-Resolved Photoelectron Spectrum (TRPES)[20,21] simulations and compared the results with new time-resolved pump/probe measurements of the ion fragment signal (reported here and also recently by Peralta Conde *et. al.*[22]), which use VUV light at 7.7 eV for both pump and probe (fifth harmonic of the Ti:Sapphire laser). This pump pulse excites the molecule almost exactly at the absorption maximum and the high energy, low intensity probe pulse leads primarily to single photon ionization which can be modeled directly, as discussed below. The simulations take into account not only the dynamics after π→π* excitation by the pump pulse, but crucially, also what happens during the ionization probe step.

## Theory

To simulate the molecular dynamics of photoexcited neutral ethylene, we used the *ab initio* multiple spawning method (AIMS), with energies, gradients and non-adiabatic coupling vectors calculated during the dynamics, i.e. "on-the-fly," at the Multi-State Second-Order Perturbation Theory level (MS-CASPT2). Briefly, we just state that the nuclear wavefunction in AIMS is composed of a linear combination of "trajectory basis functions" (TBFs) in the form of frozen Gaussian wavepackets.[23] Each TBF is associated with a single adiabatic electronic state (i.e., these are vibronic wavepackets) and its phase space center (position and momentum) evolves according to Hamilton's equations for the associated electronic state. The basis set is expanded adaptively when the nonadiabatic coupling is large, in order to describe surface crossing effects. The complex amplitudes of the vibronic wavepackets are determined during the evolution by solving the nuclear Schrödinger equation in the finite basis set. Finally, the gradients of the potential energy surfaces needed to solve Hamilton's equations and the nuclear kinetic and potential energy integrals needed to solve the Schrödinger equation are calculated during the dynamics using the MS-CASPT2 method. Further details of the AIMS and AIMS-MSPT2 method can be found in previous literature[24-26] and our recent publication,[18] respectively.

Our AIMS simulations provide a description of the time-evolving nuclear wavefunction for photoexcited neutral ethylene, and we wish to use this to calculate the TRPES signal. In principle, we could use the nuclear and electronic wavefunctions from AIMS to calculate the TRPES signal directly;[27-31] however, such an approach would require, in addition to the neutral excited-state dynamics, propagation of the nuclear wavefunction on the ethylene cation states for each pump-probe time delay. This would be computationally quite demanding and we therefore use a simplified approach, based on a method described previously.[32] This simplified approach ignores potential interferences between different outgoing channels of the ionized electron, and does not attempt to describe the angular distribution of ejected electrons.

First, we assume that the electron ejection is ultrafast so the transition is fully vertical (Franck-Condon approximation). Neglecting possible interferences involving the outgoing free electron, the instantaneous single-photon-induced ionization probability from neutral state $I$ into final cation state $\alpha$ is given as the integral over the ionization probabilities involving all possible states of the continuum electron:

$$P_{I-\alpha} = \int P_{I-\alpha\eta} d\eta \qquad (1)$$

where $\eta$ collects the quantum numbers describing the continuum electron. Within the electric dipole approximation, the ionization probability depends on the transition dipole matrix element connecting neutral and cationic states. The ionization probability vanishes when the probe pulse has insufficient energy to ionize the molecule, because there are then no allowed final states for the continuum electron. In a semiclassical limit where 1) each of the trajectory basis functions is considered independently, 2) the transition is considered to be sudden so the vibrational wavefunction does not change during the transition, and 3) the matrix element is approximated by its value at the center of the trajectory basis function (TBF), the ionization probability for a single TBF is given as:

$$P_{I-\alpha\eta}(t) \propto \left|\langle\psi_I | \vec{\varepsilon}\cdot\hat{\mu} | \psi_\alpha\phi_\eta\rangle\right|^2 \delta(\hbar\omega_{probe} - \text{IP}_{I\alpha}(R(t)) - E_{kin}(\eta)) \qquad (2)$$

where $\psi_I$ is the electronic wave function of neutral state $I$, $\psi_\alpha$ is the electronic wave function of cation state $\alpha$, $\phi_\eta$ is the wave function photoelectron orbital for quantum state $\eta$ of the ejected electron, $\vec{\varepsilon}\cdot\hat{\mu}$ is the projection of the probe polarization on the molecular dipole operator, $\hbar\omega_{probe}$ is the probe laser energy, $E_{kin}(\eta)$ is the kinetic energy of the departing electron, and $\text{IP}_{I\alpha}(R(t))$ is the vertical ionization potential at the molecular geometry $R(t)$ which is the center of the given TBF.

Within the framework of Eq. 2, the photoelectron intensity is controlled by two factors: the bound-free dipole matrix element (first term on the right hand side), which we refer to as the electronic factor, and the resonance condition (δ-function, second term on the right hand side), which we refer to as the energetic factor. The resonance condition embodied by the energetic factor enforces energy conservation before and after ionization. The bound-free dipole matrix element can be simplified if the photoelectron orbital is assumed to be orthogonal to the occupied orbitals of the neutral molecule, in which case:

$$P_{I-\alpha\eta}(t) \propto \left|\langle\phi_{Dyson}^{I-\alpha} | \vec{\varepsilon}\cdot\hat{\mu} | \phi_\eta\rangle\right|^2 \delta\left(\hbar\omega_{probe} - \text{IP}_{I\alpha}(R(t)) - E_{kin}(\eta)\right) \qquad (3)$$

where $\phi_{Dyson}^{I-\alpha}$ is a Dyson orbital, defined as an overlap integral involving the neutral and cation electronic wavefunctions:

$$\phi_{Dyson}^{I-\alpha}(\vec{r}) = \sqrt{N} \int d\vec{r}_1 \cdots d\vec{r}_{N-1} \psi_I(\vec{r}_1 \cdots \vec{r}_N) \psi_\alpha(\vec{r}_1 \cdots \vec{r}_{N-1}) \quad (4)$$

where $N$ is the number of electrons in the neutral molecule. As described previously,[32,33] we evaluate Eq. 4 using a complete active space self-consistent field (CASSCF) calculation on the neutral molecule and a CASCI (CASSCF without orbital optimization) calculation on the cation using the CASSCF orbitals determined for the neutral molecule. The remaining ingredients for the dipole matrix elements in Eq. 3 are the photoelectron orbitals. We take these to be spherical waves centered at the center of charge, in the form of a product of Coulomb-wave radial functions and spherical harmonics expanded up to L=5, *i.e.* ignoring all electron-molecule interactions beyond the asymptotic Coulomb interaction of the molecular cation. The dipole matrix elements were then evaluated on a real-space cubic grid with sides of length 10 Å and an equispaced grid with 128 grid points per side. We used the ezDyson code[33] for this purpose, which also includes isotropic orientational averaging of Eq. 3. As this is essentially a first-order Born approximation for the photoionization matrix element, we refer to these results below as BA1.

As an alternative to explicitly evaluating the matrix elements in Eq. 3, we can make use of our previous assumption[32] that at low photon energies, the probability of ionization is determined mainly from the electronic overlap of neutral and cation states (the norm of the Dyson orbital):

$$P_{I-\alpha}(E_{kin},t) \propto \left\langle \phi_{Dyson}^{I-\alpha} \middle| \phi_{Dyson}^{I-\alpha} \right\rangle \delta\left(\hbar\omega_{probe} - IP_{I\alpha}(R(t)) - E_{kin}(\eta)\right). \quad (5)$$

Indeed, such an approximation should be reasonable as the relative intensities of peaks in a photoelectron spectrum often closely follows the norm of the relevant Dyson orbital.[33] The TRPES calculated using the norm of the Dyson orbital is referred to as DN in the following.

Given the photoionization intensity for a particular nuclear geometry, evaluated either from Eqs. 1 and 3 or Eqs. 1 and 5, the total TRPES for each cation channel can be obtained as an incoherent sum over all TBFs associated with the photoexcited neutral molecule:

$$P_\alpha^{BA1}(E_{kin},t) \propto \sum_{I,i,\eta} n_i^I(t) \left| \langle \phi_{Dyson}^{I-\alpha} | \vec{\varepsilon} \cdot \hat{\mu} | \phi_\eta \rangle \right|^2 \delta(\hbar\omega_{probe} - \text{IP}_{I\alpha}(R_i^I(t)) - E_{kin}(\eta)) \tag{6}$$

or

$$P_\alpha^{DN}(E_{kin},t) \propto \sum_{I,i} n_i^I(t) \langle \phi_{Dyson}^{I-\alpha} | \phi_{Dyson}^{I-\alpha} \rangle \delta\left(\hbar\omega_{probe} - \text{IP}_{I\alpha}\left(R_i^I(t)\right) - E_{kin}(\eta)\right) \tag{7}$$

where $i$ is the index of the TBF on the $I$th electronic state, $n_i^I$ is the population associated with the TBF, and $R_i^I$ is the position center of the TBF. The results from Eqs. 6 and 7 were then convolved with a gaussian function with a full-width-half-maximum (FWHM) appropriate for the experimental time and energy resolutions.

As a final comment, the electronic wave function used in the dynamics usually cannot give equally accurate results for both the ionization potential and the excitation energy. In fact, the former is usually considerably more accurate. Therefore, the excited state ionization potential $\text{IP}_{I\alpha}$ in Eq. 6 or 7 will usually be too small. This can be corrected by introducing a constant shift $\Delta$ in $\text{IP}_{I\alpha}$ determined to ensure that electrons ionized by coincident pump/probe pulses have the correct mean kinetic energy. Specifically, $\text{IP}_{I\alpha}$ is replaced by $\text{IP}_{I\alpha} - \Delta$:

$$\Delta = V_{D_0}^{CASPT2}\left(R_{FC}\right) - V_{S_{bright}}^{CASPT2}\left(R_{FC}\right) - \text{IP}_{S_0/D_0}^{vertical,expt} + \Delta E_{S_0/S_{bright}}^{vertical,expt} \tag{8}$$

where $R_{FC}$ is the Franck-Condon point (minimum on the neutral ground electronic state), $D_0$ is the cation ground electronic state, $S_0$ is the neutral ground state, and $S_{bright}$ is the bright state (the state that the neutral molecule is excited to). By integrating the two-dimensional TRPES spectra over the photoelectron energy axis, we obtain the ion yield time trace, which corresponds to the observed lifetime in the experiments described below. Similarly, we can obtain the photoelectron spectrum at any given pump-probe time delay or, by integrating over time, the total photoelectron spectrum.

## Experiment

The apparatus is depicted in part 2 of this series and has been described in detail previously.[34,35] High-order harmonics of 807 nm are generated with a repetition rate of 10 Hz by loosely focusing (f = 6 m) 30 mJ, 50 fs laser pulses into a 5 cm gas cell with laser drilled pinholes. The cell is filled with 8.0 Torr of Ar gas and scanned through the focus to optimize the harmonic yield. The harmonic and fundamental beams are allowed to diverge for three meters where they are incident on a silicon mirror set at the 800 nm Brewster angle (75°). The silicon mirror removes the fundamental and reflects the harmonics.[36] Pump/probe delay is achieved with a split mirror interferometer (SMI) similar to that described previously.[22,37] The harmonics are focused into a pulsed molecular beam of neat ethylene by two "D-shaped" spherical concave mirrors (r = 20 cm) at normal incidence. One mirror is mounted on a piezoelectric translation stage to produce a delay. In part 2 of this series, we will present the results of experiments utilizing the 5th harmonic (hv = 7.7 eV) for the pump pulse and XUV harmonics 11-15 (hv = 17-23 eV) for the probe pulse. In this paper, we present the results of experiments using the fifth harmonic for both pump and probe pulses. The 7.7 eV photon energy lies near the maximum of ethylene's broad first absorption band which is dominated by the π→π* transition.[38,39] We selected only the 5th harmonic for pump and probe arms by inserting an interference filter (Acton Research 160-N) in both arms of the SMI. Photo-ions from the focal region are measured with a time of flight ion mass spectrometer (TOF). We checked that the results reported here are not influenced by dimers in the molecular beam by varying the backing pressure of the pulsed valve.

## Results and Discussions

The symmetric pump/probe delay ion yield signals are shown in Figure 1. The experimental finite instrument response (FIR) function (convolution of pump and probe pulses) is estimated to be a Gaussian with full width at half maximum (FWHM) of 25 ± 7 fs based on the simultaneously recorded two photon ionization of background water molecules. The extra width of the ion yields in Figure 1 is due to finite excited state lifetime after photon absorption. The $C_2H_4^+$ parent ion signal quickly decays as the nuclear wavefunction moves away from the Franck-Condon region and acquires kinetic energy, while the $C_2H_3^+$ and $C_2H_2^+$ signals persist as the nuclear wavefunction samples the excited state PES. For comparison with previous work, we fit the data with the two step model proposed by Mestdagh[13] and indicated by the multiphoton probe studies.[14,15] We get a time constant of $\tau_1$ = 21 ± 4 fs by fitting the $C_2H_4^+$ signal with a single exponential convolved with the FIR. Using the Mestdagh model, the $C_2H_3^+$ ($C_2H_2^+$) signal then gives a second time constant of $\tau_2$ = 27±5 fs (23 ± 6 fs). These short time constants are in good agreement with those using long wavelength probe pulses. However, the $\tau_2$ parameter appears slightly longer than that obtained by Peralta Conde *et. al*.[22]

We performed dynamics simulations to model the ultrafast dynamics observed in experiments. As discussed in more detail in part 2 of this series, a good understanding of the photochemical process has been achieved in terms of the conical intersections (CIs) involved, namely, the twisted-pyramidalized CI and the ethylidene-like CI, and the vibrational modes that promote the ultrafast quenching to the ground state. In terms of the excited state lifetime, the most obvious quantity to be compared directly with the experiments, the photoion signal from the electronically excited molecule (now measured in several independent experiments) decays much faster than the π→π* (V in Mulliken notation) state lifetime predicted by theory. To understand the discrepancy, we simulated the TRPES as described above. Dynamics on the neutral states were simulated as described previously.[18] Initial conditions (positions and momenta) were chosen by sampling from the Wigner distribution corresponding to the molecule in its ground vibrational state and a total of 44 such samples are included in the simulations described here. Each of the TBFs corresponding to these initial conditions were propagated independently for 200 fs, which is sufficient time that practically all the excited state population has returned to the ground electronic state. The electronic structure problem was solved using the multiconfiguration state-average complete active space with second order perturbation theory (SA-MS-CASPT2).[40] This allows for treatment of multiple electronic states and includes both static and dynamic electron correlation effects. The three lowest singlet states are included in the state averaging and the active space used has two electrons in two orbitals. The basis set used is the polarized double-zeta 6-31G* set and thus the electronic structure method can be denoted SA-3-CAS(2/2)-PT2/6-31G*.

For the electronic wavefunctions which are required to calculate the electronic factor in either the Dyson norm or first-order Born approximation, we use SA-3-CAS(6/5) and SA-5-CAS(5/5) for the neutral and cation electronic states, respectively (again with the 6-31G* basis set in both cases). The primary reason to evaluate the electronic factors using wavefunctions with a larger active space than that used in the excited state neutral dynamics was in order to describe the analogous experiment using an XUV probe pulse (where more cationic states are energetically accessible). This will be discussed in more detail in the second paper of this series, which focuses on the pump-XUV probe experiment. We note that we have checked that the results presented here (see below) are not sensitive to this choice and practically identical results are obtained.

The vertical ionization potential was calculated with the analogous second-order perturbation theory corrected method at each of the molecular geometries. As discussed above, the computed vertical ionization potential was shifted according to Eq. 8, giving rise to a Δ value of 1.61 eV. In detail, the calculated IP and $S_1$ excitation energy are 10.22 eV and 8.98 eV at the $S_0$ minimum geometry (Franck-Condon point), while the experimental results are 10.51 eV[41] and 7.66 eV[42], respectively. After excitation, the ionization cross sections were calculated every 10 fs with a probe photon energy of 7.7 eV for 200 fs along the TBFs from the dynamics of the excited neutral molecule. The spectrum from these calculations was generated and then convolved with Gaussian instrument response functions in time (25 fs FWHM) and energy (0.1 eV FWHM) to simulate the experimental results.

For comparison with the experiment that measures only the ion yield and not the time-resolved photoelectron energy distribution, we integrated over the photoelectron energy variable of the TRPES spectrum to obtain the time trace of the signal. Since the probe pulse always follows the pump pulse in our simulations, we add the calculated signal to its mirror image (reflected through $t$=0) to represent the signal from negative time delays. The integrated TRPES signals and the total measured photoion yield (scaled to unit maximum) are plotted in Figure 2. Both the first-order Born approximation (BA1) and Dyson-Norm (DN) methods show excellent agreement with the experimental data, although the BA1 method is perhaps slightly better. A similar analysis can be done under the assumption that all excited state population gives rise to an ion signal, i.e. that the molecule is always ionizable before it returns to the ground electronic state $S_0$. The resulting expected ion yield is plotted in Figure 2 (pink line), subject to the same reflection across the time axis to account for pump/probe and probe/pump signals. As could be expected from the 89 fs excited state lifetime predicted by our previous AIMS-MSPT2 calculations, the photoion yield decays faster than the predicted excited state population. We take this as a clear indication that it is questionable to assume that the molecule is always ionizable when it is in the excited electronic state.

Two factors are obvious targets for understanding the discrepancy between the excited state population decay and the decay of the photoion yield, namely the energetic factor and the electronic factor in Eqs. 6 or 7. To investigate their contributions to the signal loss, we carried out TRPES calculations under different conditions. The results are shown in Figure 3. In all cases, these spectra are normalized to the same area, as the total photoelectron yield is not predicted by Eqs. 6 or 7 (this would require much more detailed calculations including potential interferences between outgoing electron channels.) The upper left panel (Figure 3A) shows the predicted TRPES using the experimental 7.7 eV probe energy and the Dyson-Norm (DN) method of Eq. 7. The upper right panel (Figure 3B) shows the predicted TRPES using the 7.7 eV probe energy and the first-order Born approximation (BA1) method of Eq. 6. While the photoion yield from the DN and BA1 methods is quite similar (see Figure 2), the TRPES are noticeably different. Specifically, there are fewer high energy (above 2 eV) photoelectrons in the BA1 spectrum. Nevertheless, the BA1 and DN spectra differ by little more than a constant scaling, in accord with the experimental observation that at relatively low probe energy, the photoionization cross section is only weakly dependent on the kinetic energy of the departing electron.[43,44] This is of course the assumption which leads from the BA1 to the DN methods. In the lower left panel (Figure 3C), we show the TRPES which results if the probe energy is increased by 1.61 eV (to 9.3 eV). This value remains below the ground state ionization potential and therefore, in principle, represents a feasible experiment (ionization only occurs if the molecule has first been excited by the pump pulse). With the increased energy of the probe photon, the calculations now predict a longer observed photoelectron (photoion) lifetime. This shows that the observed photoion lifetime is in part shortened by an energetic condition, i.e. the ionization potential of the excited ethylene molecule can be larger than the 7.7 eV provided by the probe photon. In fact, integrating this TRPES over all photoelectron energies gives a predicted photoion signal which decays on the same timescale as the $S_1$ population (89 fs) reported in our previous work.[18] Finally, in the lower right panel (Figure 3D), we show the TRPES calculated by assuming that the electronic factor (i.e. the Dyson orbital norm in Eq. 7) is always unity and using the experimental 7.7 eV probe photon energy. Comparison with the analogous DN TRPES in Figure 3A shows that although the electronic factor does have a role in shaping the observed signal, it does not lead to a large change in the predicted lifetime (extent of the TRPES spectrum along the time axis). Thus, we conclude that the energetic factor plays the major role in the observed lifetime from both the current and previous time-resolved photoionization experiments on ethylene. This was previously suggested both by our group[45] and also subsequently by Barbatti, *et. al.*[46] However, this is the first direct demonstration of the validity of this suggestion. It is also the first prediction of the full TRPES as shown in Figure 3. These may be measured in the future and it will be especially interesting to see these at both 7.7 eV and 9.3 eV probe photon energies.

To better demonstrate the energetic effect in the TRPES, we plot the photoelectron energies (for the 7.7 eV probe photon) at the center of each ionizable TBF during the excited neutral dynamics (blue dots) along with the percentage of the TBFs on the excited state ($S_1$) that are ionizable (black line) in Figure 4. Both the photoelectron energies and the percentage of the TBFs that can be ionized decrease as the time delay between the pump and probe pulses increases on the excited state.

There are indications from previous experiments that isotope effects are expected to be minor in the dynamics of excited ethylene.[14,15] Thus, we carried out similar AIMS-MSPT2 calculations for fully deuterated ethylene ($d_4$-$C_2H_4$). The VUV/VUV pump-probe experiments described here have not yet been performed on deuterated ethylene, so there is no experimental data for comparison (the previous experiments on $d_4$-$C_2H_4$ have used multiphoton ionization for the probe step). Thus, we compare the time trace (integrating over all photoelectron energies) of the TRPES spectra using the BA1 method of Eq. 6 for normal and deuterated ethylene in Figure 5. As can be seen, the predicted photoion yield does not depend sensitively on isotopic substitution. This is a consequence of the fact that the early dynamics is dominated by C-C stretch and torsional motion, neither of which is strongly affected by isotopic substitution. A small difference between the normal and deuterated ion yields is seen for pump-probe time delays exceeding 100 fs. This arises because some portion of the excited state population undergoes hydrogen migration to form ethylidene at later times. However, this is a minor channel and has a negligible effect on the excited state lifetime, as discussed previously[18] and in more detail in part 2 of this series.

## Conclusion

We have employed a novel VUV/XUV pump-probe apparatus and AIMS simulations (at the MS-CASPT2 level, including both static and dynamic electron correlation effects) to study the dynamics of the prototypical ethylene molecule upon $\pi \rightarrow \pi^*$ excitation. In this first part of a two-part series, we have elucidated the origin of the discrepancy of the excited state lifetime between ion yield measurements and theoretical simulations. The energetic factor representing the instantaneous ionization potential of the excited neutral molecules dominates the fast signal loss observed in experiment. Somewhat surprisingly, even a 7.7 eV probe photon is insufficiently energetic to ionize ethylene for its entire sojourn on the excited state. Thus, the observed lifetime for the photoion yield is significantly shorter than the predicted excited state lifetime. We provide predictions for the time-resolved photoelectron spectra of ethylene, both at the 7.7 eV probe photon energy used in the experiments we describe here and also at a larger 9.3 eV probe photon energy which is predicted to lead to a photoion yield decay that is more in line with the excited state lifetime.

The Rydberg states of ethylene were not included in this study. The strong agreement between theory and experiment obtained here suggests that the presence of Rydberg states energetically in the vicinity of the $\pi\pi^*$ state does not dramatically alter the picture of nonradiative decay of ethylene. This is indeed verified by our dynamics simulation with Rydberg states (unpublished, Mori *et al.*). As all molecules absorb in the VUV and XUV, the experimental techniques used here are expected to be widely applicable.

## Acknowledgements

The theory work was performed under DOE Contract No. DE-AC02-7600515. The experiment was supported by the US Dept. of Energy Office of Basic Energy Sciences, under contracts numbers DE-AC02-05CH1123 and DE-FG-52-06NA26212. We acknowledge W. G. Glover, C. R. Evenhuis and T. Mori for helpful discussion. A.M. Stooke gratefully acknowledges the full support of the Fannie and John Hertz Foundation. We thank A. Stolow for helpful discussions. We acknowledge C. Caleman, M. Bergh, H. Merdji, and M. P. Hertlein for help with the apparatus in its early stages.

**Figures and Captions**

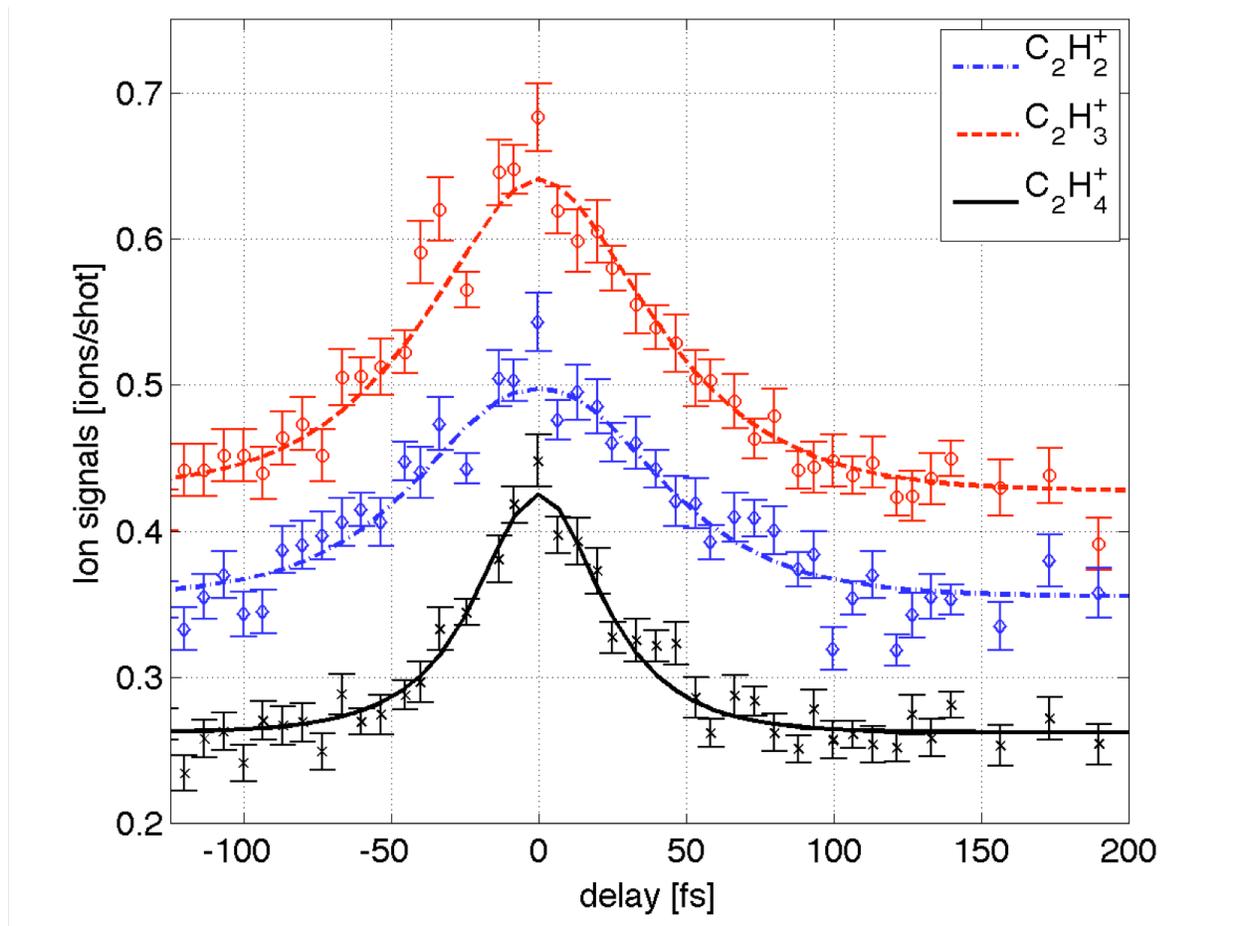

**Figure 1.** Time resolved photoion yield from 7.7 eV pump/7.7 eV probe experiments. The $C_2H_4^+$ signal (black x's) is modeled with a single exponential decay with a time constant of 21 fs convolved with the finite instrument response (solid black curve). The $C_2H_3^+$ (red circles) and $C_2H_2^+$ (blue diamonds) signals are modeled with a two-step exponential decay model described in the text.

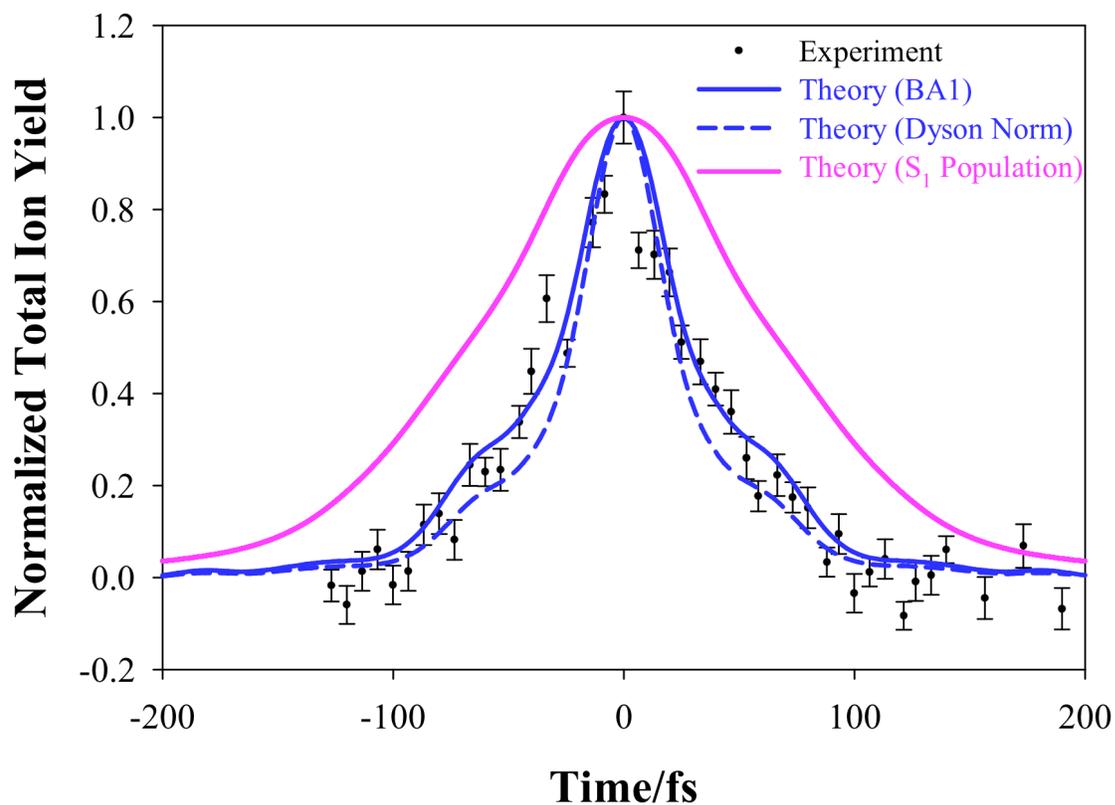

**Figure 2.** Comparison between experimental total ion yield and AIMS-MSPT2 predicted signals. Experimental signal is shown as black dots with error bars. The calculated signals (from integrating the AIMS-MSPT2 predicted single-photon TRPES spectra over all photoelectron energies) are shown in blue lines. The solid blue line is the first-order Born approximation (BA1) method of Eq. 6 and the dashed blue line is the Dyson-Norm (DN) method of Eq. 7. The pink line shows the photoion yield that would result from assuming that all $S_1$ population is ionizable. This assumption leads to much slower decay of the photoion signal compared to the experiment.

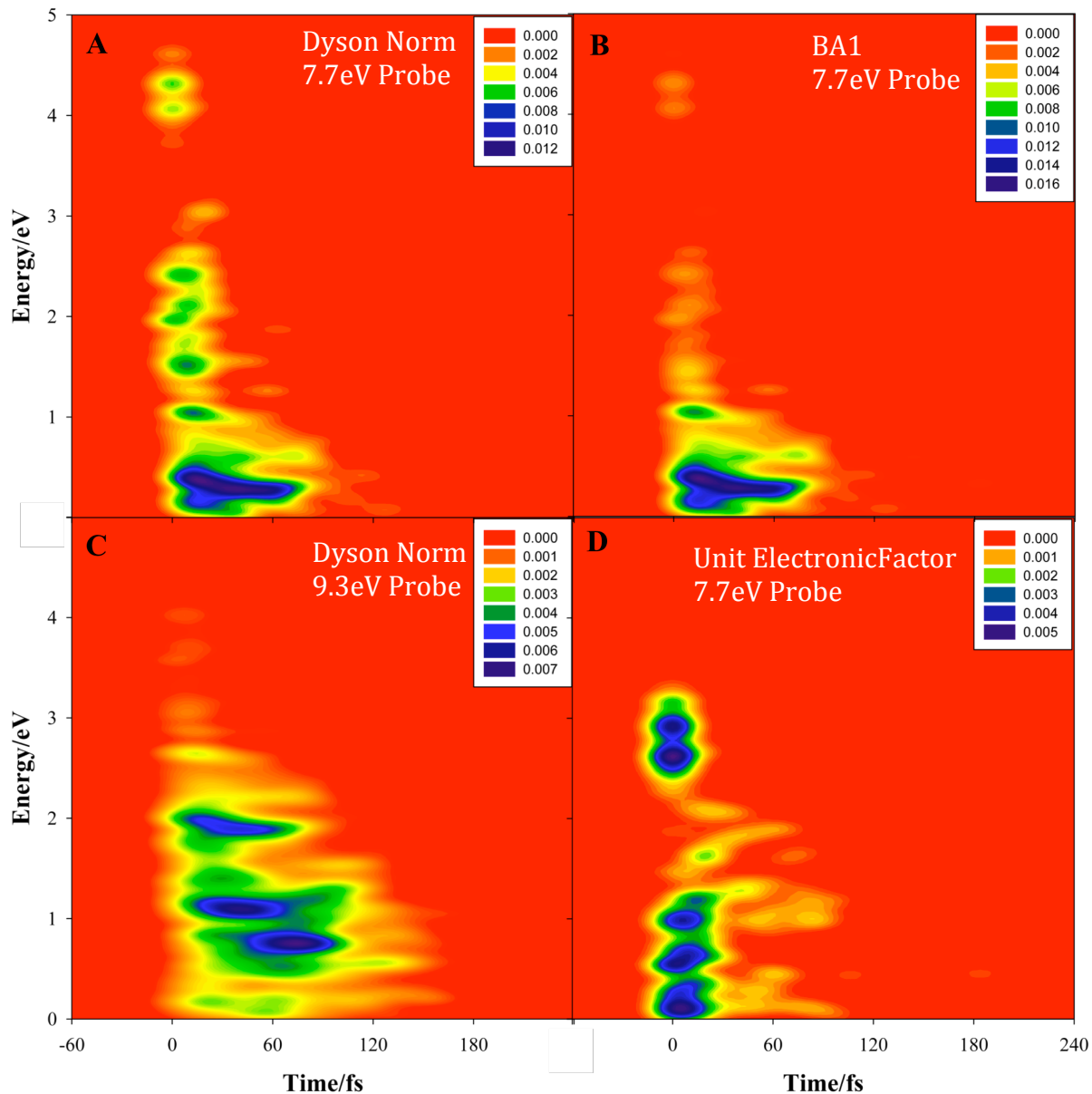

**Figure 3.** Calculated time-resolved photoelectron spectra (TRPES) under different conditions. A) Dyson-Norm (DN) method with 7.7 eV probe. B) First-order Born approximation (BA1) with 7.7 eV probe. C) DN method with 9.3 eV probe. D) TRPES calculated with 7.7 eV probe where the electronic factor is set to unity. Comparison of A and B shows that the difference between the BA1 and DN methods is more visible in the TRPES than in the photoion yield (see Figure 3). Comparison of B and D shows that most of the photoelectron (and therefore photoion) yield decay is due to the energetic factor (population which is not ionizable) and not the electronic factor. Panel C shows that TRPES with a 9.3 eV probe is predicted to yield a slower observed photoelectron decay.

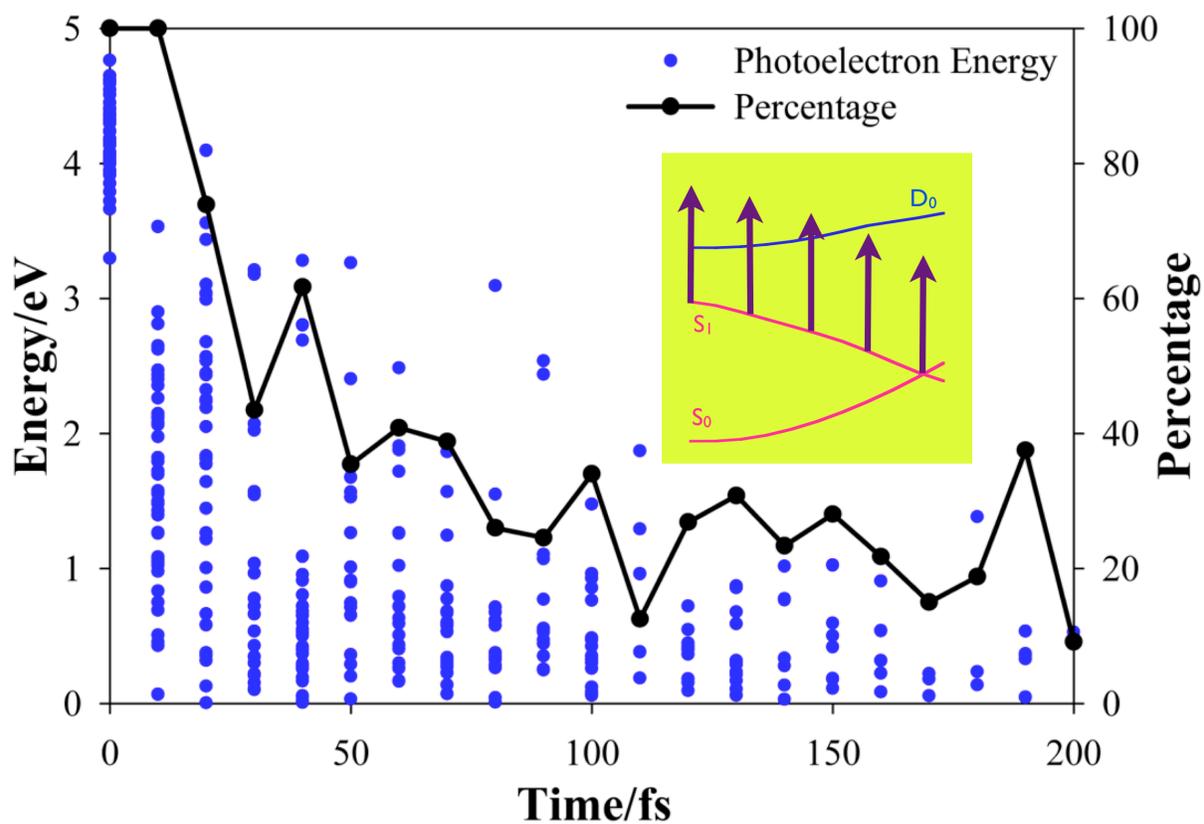

**Figure 4.** Analysis of the energetic factor during the neutral excited state dynamics. The blue dots show the energies of departing photoelectrons (for 7.7 eV probe photon energy). The black lines show the percentage of TBFs on the excited state ($S_1$) which can be ionized for the given pump-probe time delay. Both the photoelectron energy and the percentage of TBFs that can be ionized are decreasing as the pump-probe time delay increases. The inset sketches the potential energy surfaces of the neutral and cation states to schematically indicate the physical picture, where the vertical arrow represents the energy of the probe photon.

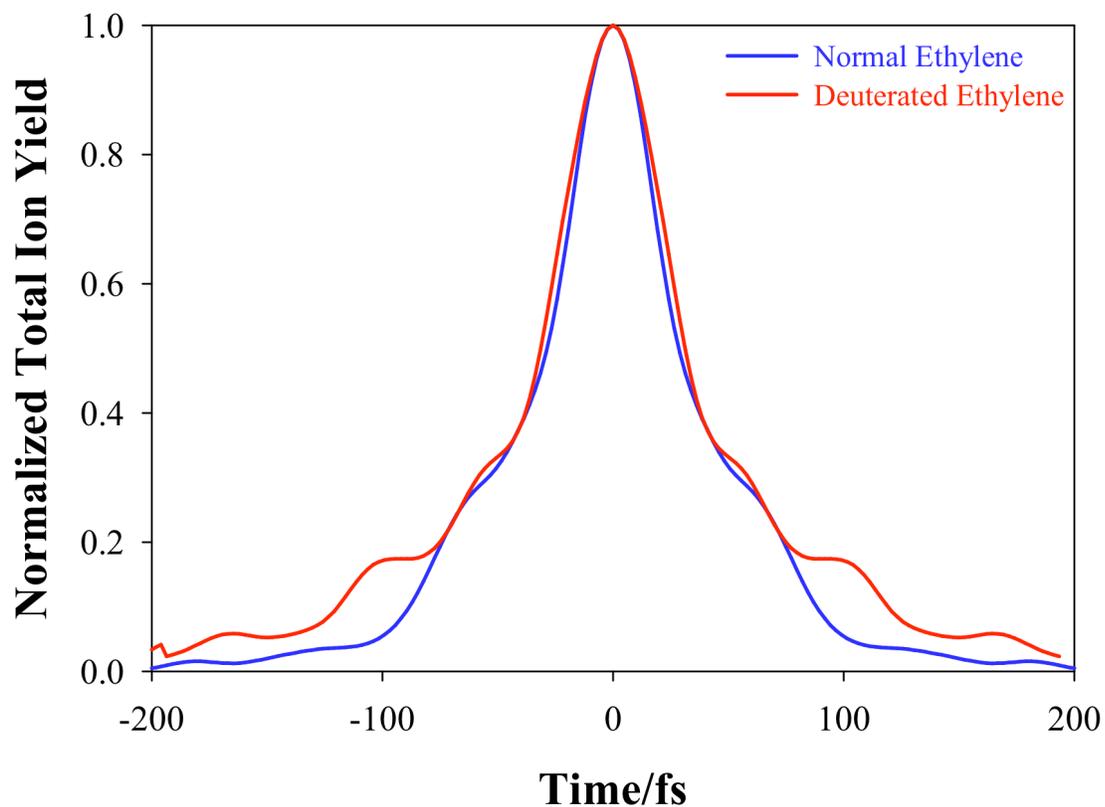

**Figure 5.** Comparison between calculated ion yield signals for fully deuterated ethylene ($C_2D_4$) and normal ethylene ($C_2H_4$). No isotope effect is observed until pump-probe time delays of ~100fs, when the few remaining excited state neutral molecules begin to access the ethylidene-like CI region.